\input{aipcheck}

\documentclass[final
  ]
  {aipproc}

\layoutstyle{8x11single}

\begin{document}

\title{Current and Future Liquid Argon Neutrino Experiments}


\author{Georgia S. Karagiorgi}{
  address={Department of Physics, Columbia University, New York, NY 10027, USA}
}

\begin{abstract}
The liquid argon time projection chamber (LArTPC) detector technology provides an
opportunity for precision neutrino oscillation measurements, neutrino cross section measurements, and searches for rare processes,
such as SuperNova neutrino detection. These proceedings review current and future LArTPC neutrino experiments. 
Particular focus is paid to the ICARUS, MicroBooNE, LAr1, 2-LArTPC at CERN-SPS, LBNE, and 100~kton at Okinoshima experiments. 
\end{abstract}

\maketitle


\section{Liquid Argon Time Projection Chamber Experiments}

Over the last decade, the neutrino community has demonstrated a tremendous amount of interest in the use of
liquid argon time projection chambers (LArTPC's) as detectors for neutrino experiments.
LArTPC's rely on measuring the ionization energy loss of any charged particle, $dE/dx$, 
in order to reconstruct the particle's trajectory, momentum, range, and, consequently, type.
In the case of LArTPC neutrino detectors, a large, high-purity liquid argon volume serves as both the neutrino interaction medium and the
tracking medium for charged particles produced in neutrino interactions. Ionization charge produced along 
charged particle tracks drifts toward one side of the detector, under the influence of an electric field applied uniformly 
within the liquid argon volume, and then gets collected on a set of finely segmented ``sensor-wire planes''.
Precise charge amplitude(s) vs.~wire position(s) and arrival time(s) are recorded for all collected charge, and used to reconstruct the event topology.
Argon scintillation light emitted during the event is also
typically detected by photo-sensitive detectors (typically, PMT's), and it is used for extracting the initial interaction time, $t_0$, as well as for event triggering.

\subsection{Motivation and challenges}
\label{challenges}

Liquid argon is dense and relatively inexpensive; this makes it an ideal detector medium for
low rate TPC's, which are applicable to the study of neutrino interactions. 
LArTPC detectors offer high detection efficiency as well
as higher background rejection power relative to the most commonly used neutrino detection technologies:
Cherenkov and scintillator detectors. 
Typical LArTPC detector performance, demonstrated by the ICARUS and ArgoNeuT LArTPC experiments, and elsewhere
\cite{Rubbia:2011ft,Antonello:2012hu,Anderson:2012vc,t32}, 
corresponds to mm-scale spatial 3D resolution, 3\%/$\sqrt{E\mathrm{(GeV)}}$ electromagnetic shower energy resolution, 
and $>90$\% electron vs.~photon differentiation. Electron vs.~photon differentiation is achieved
by measuring the early-on ionization from a single electron or from the $e^+e^-$ pair produced from a photon which 
converts on an argon nucleus. The single electron early-$dE/dx$ typically corresponds to that of one minimum ionizing particle (MIP), while that of the $e^+e^-$ pair typically corresponds to that of two MIP's. 
The high electron/photon differentiation in particular makes LArTPC's an ideal
technology for $\nu_e$ measurements. 
The high performance of LArTPC's is linked to their high ionization charge yield ($\sim1$~fC/mm for MIP's) 
and small charge diffusion ($\sim$~mm for several meters of drift). Additional advantages include the fact that the detector volume is
homogeneous and fully active, which simplifies event reconstruction and maximizes detection efficiency. The high
scintillation yield of liquid argon also provides an attractive triggering mechanism.

Despite the attractiveness of this technology, there are several challenges that need to be overcome before scaling LArTPC detectors
to sizes sufficiently large (multi-kiloton scale) for high-statistics and, therefore, high-precision neutrino measurements. Such challenges include the development and successful operation of large cryogenic systems; the demonstration of long ionization charge drift distances, which requires ultra high purity when evacuation prior to filling is impractical, and high voltage (on the order of 100~kV or more) on the TPC cathode; the development of readout and data acquisition systems which can handle, process, and store data from a large number of readout channels (many tens of thousands for a kiloton scale detector) sampled at MHz rates over long time periods dictated by charge drift distances (on the order of many milliseconds per event); the development of low noise, cold electronics; and the development of fast reconstruction software. Those challenges are being addressed by a number of ongoing and planned R\&D projects, including ArgoNeuT, ICARUS, the Materials Test Stand and LAPD at Fermilab \cite{Rebel:2011zzb}, LArIAT \cite{lariat2}, MicroBooNE \cite{Chen:2007ae}, ArgonTube at Bern \cite{Badhrees:2012zz}, the 50~L and 10~m$^3$ demonstrators at CERN \cite{Arneodo:1998ef,Arneodo:2003vh}, and the Test-Beam (T32) experiment at J-PARC \cite{Araoka:2011pw}.

\subsection{Physics goals}

LArTPC neutrino experiments aim to address outstanding fundamental questions in neutrino physics. Those include the search of
CP violation in three-neutrino oscillations, through its observable effects in long-baseline $\nu_e$ and $\bar{\nu}_e$ appearance,
and the determination of the neutrino mass hierarchy. The LBNE \cite{lbne}, LAGUNA/LBNO \cite{Kisiel:2009zz}, 
100-kton at Okinoshima \cite{Badertscher:2008bp}, MODULAr \cite{Baibussinov:2007ea}, and GLADE \cite{Thomas:2012zzb} experiments aim 
to search for and possibly measure CP violation, as well as measure the neutrino mass hierarchy. Due to their detector sizes, those experiments also provide opportunities for searches for rare events, including proton decay and baryon number violating processes, SuperNova core collapse neutrinos, and, potentially diffuse SuperNova neutrino background detection \cite{Cavanna}. Two equally fundamental but also experimentally pressing questions are (1) whether sterile neutrino oscillations
take place at short baselines, and (2) whether and how well we understand inclusive and exclusive neutrino cross 
sections, and in particular nuclear effects and final state interactions in neutrino-nucleus scattering. The former question forms
the main physics goal of the LAr1 \cite{Chen:2012nv} and the 2-LArTPC at CERN-SPS \cite{Antonello:2012qx} experiments, 
and one of the goals of the MicroBooNE experiment.
The latter question is being addressed by current and near-term LArTPC experiments including ArgoNeuT, MicroBooNE, LAr1, 2-LArTPC at CERN-SPS,
and ICARUS.

The goal of next-generation neutrino cross-section experiments, including the ones above, is to unambiguously measure neutrino cross sections at $\sim1$~GeV. Past cross section measurements (from K2K, MiniBooNE, SciBooNE, MINOS, NOMAD) have revealed limitations in our understanding of neutrino interactions (see, e.g.~\cite{katori}). This limitation is partly a consequence of the leptonic system providing an incomplete description of neutrino interaction processes. It is now generally accepted that hadronic effects play a critical role and should be considered as well. More specifically, a precise measurement of the hadronic system (vertex activity, hadronic final state multiplicity and momentum, etc.) in neutrino interactions is necessary, and it provides critical information for testing existing models and developing more robust neutrino interaction event generators for oscillation and other neutrino physics. LArTPC's can study\footnote{One of the limitations of LArTPC's from the perspective of cross section physics is that only one type of target nucleus (Ar) is available; there are no free proton targets. On the other hand, this provides a clean measurement of nuclear effects in a single nucleus, where effects from free and bound nucleons are not convoluted (unlike the case of CH2 in MiniBooNE, for example). Another limitation is that sign identification (neutrino vs.~antineutrino) is difficult on an event-by-event basis. It is worth noting that magnetized LArTPC's are challenging. Alternative options include the use of high-purity sign-selected neutrino beam, the use of a spectrometer together with the LArTPC for lepton ($\mu$) charge identification, as in the case of the ArgoNeuT and 2-LArTPC at CERN-SPS experiments, or the use of the Michel electron from muon decay as a tag, as 76\% of $\mu^-$ capture in Ar.} events after final state interactions and provide measurements of the hadronic system in exquisite detail. Particular interaction channels can be studied in terms of final state multiplicity, as well as in terms of reconstructed neutrino energy from lepton kinematics alone vs.~summed total energy. Resulting measurements should help constrain generator-level implementations of neutrino cross section and nuclear effects models, which seem to predict vastly different final state multiplicities \cite{Golan}. Channels of interest which can be studied in current and near-future LArTPC detectors include: (1) neutrino-nucleon charged-current (CC) quasi-elastic scattering; (2) neutrino-nucleon neutral-current (NC) elastic scattering; (2) Kaon production; (3) single-pion production; (4) hyperon production; (5) single-photon production in low-energy scattering; (6) $\nu_e$ CC scattering (first high statistics measurements at $\sim$1~GeV).

\subsection{Worldwide efforts}

There is a significant worldwide effort to develop the LArTPC technology to multi-kiloton detectors for future neutrino experiments, both in the form
of dedicated R\&D and test experiments such as those mentioned above, and in the form of intermediate-scale (multi-ton scale) LArTPC experiments.
Table~\ref{tab:a} summarizes present and future LArTPC experiments. Selected current and future LArTPC experiments are presented in the following two sections.

\begin{table}
\begin{tabular}{lclcclll}
\hline
    \tablehead{1}{l}{b}{Experiment\\}
  & \tablehead{1}{c}{b}{LAr mass\\(tons)}
  & \tablehead{1}{l}{b}{Physics\\goals}
  & \tablehead{1}{c}{b}{Baseline\\(km)}
  & \tablehead{1}{c}{b}{$E_{\nu}$\\(GeV)}
  & \tablehead{1}{l}{b}{Detector\\location}
  & \tablehead{1}{l}{b}{Current\\status}
  & \tablehead{1}{l}{b}{Online\\} \\
\hline
ICARUS & 600 & R\&D, long & 732 & $\sim5-25$ & Gran Sasso & Running & Fully \\
  & & baseline (single & & & (CNGS & & operational \\
  & & detector) & & & beam) & & in 2010 \\
\hline
ArgoNeuT & 175$\ell$ & R\&D, cross & 1 & $\sim0.1-10$ & NuMI & Completed & N/A \\
 & & sections & & & & & \\
\hline
MicroBooNE & 170 (86 & R\&D, short & 0.470 & $\sim0.1-3$ & Fermilab & Under & 2014 \\
  & active) & baseline (single & & & (BNB) & construction  & \\
  & & detector) & & & & & \\
\hline
LAr1 & $17+60$ & Short baseline & $0.100$ & $\sim0.1-3$ & Fermilab & Letter of & $\sim5$~yrs \\
     & $+1000$ & (3 detectors) & $+0.470$ & & (BNB) & Intent & \\
     & (fiducial) & & $+0.700$ & & & & \\
\hline
2-LArTPC at & $150 + 478$ & Short baseline & $0.3 + 1.6$ & $\sim2$ & CERN & Proposal & $\sim5$~yrs \\
  CERN-SPS & (fiducial) & (2 detectors) & & & (new beam  & & \\
 & & & & & from SPS) & & \\
\hline
MODULAr & 5,000 & Long baseline & 730 & $\sim5-25$ & Gran Sasso & Planned & $5-10$~yrs \\
  & & (shallow depth) & & & & & \\
\hline
GLADE & 5,000 & Long baseline & 810 & $\sim0.5-2$ & NuMI & Letter of & $5-10$~yrs \\
  & & (surface) & & & off-axis & Intent & \\
\hline
LBNE & 35,000 & Long baseline & 1300 & $\sim0.5-5$ & SURF & Planned & 10+ yrs \\
  & (fiducial) & (smaller, surface & & & (new & (CD-1) & \\
  & & far detector & & & Fermilab & & \\
  & & initially) & & & beam) & & \\
\hline
LAGUNA/ & Start with & Long baseline & 2300 & $\sim$few & Europe & Expression & 10+ yrs \\
  LBNO & 20,000 & (underground & & & (new CERN & of Interest in & \\
  & & far detector) & & & beam) & preparation & \\
\hline
100-kton at & Up to & Long baseline & 665 & $\sim0.5-2$ & Okinoshima & R\&D & 10+ yrs \\
  Okinoshima & 100,000 & (underground & & & (new J-PARC & Proposal at & \\
  & & far detector) & & & beam) & J-PARC & \\
\hline
\end{tabular}
\caption{Summary of present and future LArTPC neutrino experiments. For a detailed description of the ArgoNeuT experiment and preliminary cross section results, see~\cite{Palamara,Partyka}.}
\label{tab:a}
\end{table}

\section{Current Experiments}

\begin{enumerate}
\item{{\it ICARUS:} The ICARUS experiment is widely considered as the pioneer LArTPC neutrino experiment. The detector is located underground at Gran Sasso National Lab, Italy, and detects neutrinos from the CNGS beam from CERN. The detector consists of two identical modules, $3.6\times3.9\times19.6$~m$^3$ each, and each houses two TPC's, with 1.5~m drift length. The total (active) liquid argon mass is 600 (476) tons. The detector is read out by 54,000 wire channels, with triggering provided by PMT's. The ICARUS detector performance is documented in \cite{Rubbia:2011ft}. 
The neutrino baseline is 732~km, and the neutrino mean energy is $\sim17$~GeV. ICARUS has been fully operational and collecting data since 2010, and has thus far collected data for a total of $\sim5$E19 protons on target (POT). More than half of the data (3.3E19 POT) has been analyzed so far. 
Physics results from ICARUS include: a search for sterile neutrino oscillations \cite{Antonello:2012pq}; a search for superluminal neutrinos through neutrino Cherenkov-analogue radiation \cite{ICARUS:2011aa}; and a precision measurement of the neutrino time-of-flight with the CNGS beam \cite{Antonello:2012be,Antonello:2012hg}. The CNGS events analysis is ongoing.}

\item{{\it MicroBooNE:} The MicroBooNE experiment is currently under construction. It is scheduled to begin its operation in the Fermilab Booster Neutrino Beamline (BNB) in 2014, where it will collect data for a total of 6.6E20 POT, in neutrino running mode. The neutrino baseline is 470~m, and the mean neutrino energy is a few hundred MeV. The MicroBooNE cryostat holds 170 tons (86 active tons) of liquid argon. The TPC dimensions are $2.5\times2.3\times10.2$~m, with 2.5~m being the drift direction. MicroBooNE uses three wire planes, with 3~mm wire separation (8256 wires total) and 30 PMT's which provide the event $t_0$ and triggering information. One of the primary goals of MicroBooNE is the investigation of the nature of the $\nu_e$-like excess previously observed by MiniBooNE, a Cherenkov experiment \cite{AguilarArevalo:2008rc}. The MiniBooNE excess appears at low reconstructed neutrino energy, reconstructed from lepton kinematics assuming quasi-elastic (QE) scattering. It is consistent with event signatures with either a single observable electron (e.g.~$\nu_e$ CCQE events), or a single observable photon (e.g.~NC backgrounds from $\Delta$ production and radiative decay). Because of its high electron/photon separation capability, MicroBooNE will be able to determine if the excess is due to single electron or due to single photon events, with 5.7$\sigma$ or $4.1\sigma$ statistical significance, respectively. In the case of a single electron excess, possible explanations include non-standard $\nu_{\mu}\rightarrow\nu_e$ oscillations (e.g.~due to sterile neutrinos). In the case of a single photon excess, possible explanations include new mechanisms of single photon production. The second primary goal of MicroBooNE is to provide high-statistics neutrino exclusive final state and cross section measurements in the 1~GeV neutrino energy range. MicroBooNE expects to observe a total of $\sim100$k  ($\sim30$k) $\nu_{\mu}$ CC (NC) events during its full neutrino run in the BNB, as well as a significant number of $\nu_e$ CC events ($\sim2,000$/yr) from the off-axis, higher-energy NuMI beam from Fermilab. Other physics goals include the study of backgrounds to proton decay and baryon number violating processes for larger (underground) detectors, and potential detection of SuperNova neutrinos.}

\end{enumerate}

\section{Future Experiments}

\begin{enumerate}

\item{{\it LAr1:} The LAr1 experiment is a proposed experiment which is based on continued MicroBooNE running in the Fermilab BNB, with the addition of a second, larger LArTPC detector at 700~m, and possibly a third, smaller LArTPC detector at 100~m. The smaller and larger detector fiducial masses are 17 and 1000 tons, respectively. The 1~kton detector, from which LAr1 gets its name, serves as an engineering prototype for LBNE, as it employs the ``membrane cryostat'' design of the LBNE far detector. Following that design, the TPC is constructed as an array of modular units of ``anode plane assemblies'', $2.7\times7\times0.10$~m$^3$, and ``cathode plane assemblies'', $2.5\times7$~m$^2$. 
With such detector configuration, a near/far comparison of detected BNB neutrino event spectra allows for a definitive test of the sterile neutrino oscillation hypothesis. Preliminary sensitivity estimates show coverage of the LSND 99\% and 90\% CL allowed two-neutrino oscillation parameters at $5\sigma$ in neutrino and antineutrino running mode, respectively. The LAr1 collaboration has submitted a Letter of Intent to the Fermilab Directorate, and there is strong ongoing effort to develop this idea into a proposal by Summer 2013. The projected start of construction, if the experiment is approved, is $\sim$2016.}

\item{{\it 2-LArTPC at CERN-SPS:} An analogous effort for a short baseline sterile neutrino oscillation search is being proposed in Europe, which involves the construction of two or three (optional) LArTPC detectors situated on-axis in a newly proposed neutrino beam from the CERN-SPS, with mean neutrino energy of $\sim2$~GeV. Two LArTPC detectors together with one iron spectrometer per detector will be situated at 330~m (near) and 1600~m (far) from neutrino production. A third LArTPC+spectrometer detector is being considered for a mid (1100~m) position in the same beamline. The ICARUS detector could be transported to CERN and used as the far detector, while a new 150~ton LArTPC would serve as the near detector. The two-detector configuration alone can exclude the LSND 99\% CL allowed region at 99\% CL in both neutrino and antineutrino running mode, after one and two years of running in each mode, respectively.}

\item{{\it LBNE:} LBNE is a planned long-baseline neutrino experiment, aiming to search for and measure leptonic CP violation and to determine the neutrino mass hierarchy. It consists of a near detector at Fermilab, and an underground, 35~kton (fiducial) LArTPC far detector at SURF. The experiment uses a new, high-power (2.3~MW), broadband, sign-selected $\nu_{\mu}$ beam from Fermilab, with $E_{\nu}\sim0.5-5$~GeV. The far detector baseline is 1300~km. In addition to long-baseline oscillation physics, LBNE also has sensitivity to non-accelerator neutrino measurements (atmospheric and SuperNova neutrinos) and rare processes such as proton decay. In 2012, the LBNE experiment adopted the recommendation for a ``phased approach'' toward its final design and physics goals \cite{lbnewhitepaper}. The goal of the first phase is a 2.3~MW-capable broadband beam, a $>10$~kton underground far detector, and a near detector. The current funding guidance from the U.S.~DOE allows for construction of the beamline, beam monitors, and a large LArTPC on-surface at SURF. Underground placement, the near detector, and some beamline enhancements are expected to be enabled by non-DOE sources. LBNE was granted CD-1 approval by the U.S.~DOE in 2012.}

\item{{\it 100-kton at Okinoshima:} Analogous to the LBNE effort in the U.S. is the proposal for a 100 kton LArTPC far detector at Okinoshima Island, Japan, combined with a new, high-intensity neutrino beam from J-PARC. The neutrino beam energy corresponds to $\sim$1~GeV, and the neutrino baseline is 660~km, with the detector situated at 0.76 degrees off-axis. The detector makes use of the GLACIER design, which uses a double-phase ``LEM-TPC''. This concept makes use of ionization charge drifting from liquid to gas argon phase to provide additional charge amplification before extraction by wire planes in the gas phase. This concept leads to an improved signal to noise ratio ($S/N\sim100$) compared to single-phase LArTPC operation ($S/N\sim15-30$). The proposed plan is 5 years of neutrino mode running, possibly supplemented with an additional 5 years of antineutrino mode running. The physics goals of the experiment are similar to those of LBNE. An R\&D proposal has been submitted to J-PARC.}

\end{enumerate}

\section{Conclusions}
The LArTPC technology is maturing and has become a credible alternative to conventional neutrino detectors. LArTPC detectors offer unique and superior imaging performance, and can therefore be used for physics measurements where excellent energy resolution and good background rejection are required. The neutrino community has already begun to see examples of cross section and oscillation measurements from this new technology, and much more results are expected over the next decade. New ArgoNeuT and ICARUS results are expected over the next 2-3 years, followed by results from MicroBooNE, which begins data taking in 2014. Experiments which may begin construction over the next 5-10 years, if approved, include LAr1, 2-LArTPC at CERN-SPS, GLADE, and MODULAr. Experiments on a 10+ year timescale include LBNE, LAGUNA/LBNO, and the 100-kton at Okinoshima.



\bibliographystyle{aipproc}

\bibliography{karagiorgi-nuint2012-arx}

\end{document}